# Statistical mechanical foundations of power-law distributions[*)]

A. K. Rajagopal[1] and Sumiyoshi Abe[2]

[1]*Naval Research Laboratory, Washington D. C. 20375-5320, USA*
[2]*Institute of Physics, University of Tsukuba, Ibaraki 305-8571, Japan*

**ABSTRACT**

The foundations of the Boltzmann-Gibbs (BG) distributions for describing equilibrium statistical mechanics of systems are examined. Broadly, they fall into: (i) probabilistic approaches based on the principle of equal *a priori* probability (counting technique and method of steepest descents), law of large numbers, or the state density considerations and (ii) a variational scheme - maximum entropy principle (due to Gibbs and Jaynes) subject to certain constraints. A minimum set of requirements on each of these methods are briefly pointed out: in the first approach, the function space and the counting algorithm while in the second, "additivity" property of the entropy with respect to the composition of statistically independent systems. In the past few decades, a large number of systems, which are not necessarily in thermodynamic equilibrium (such as glasses, for example), have been found to display power-law distributions, which are not describable by the above-mentioned methods. In this paper, parallels to all the inquiries underlying the BG program described above are given in a brief form. In particular, in the probabilistic derivations, one employs a different function space and one gives up "additivity" in the variational scheme with a different form for the entropy. The requirement of stability makes the entropy choice to be that proposed by Tsallis. From this a generalized thermodynamics description of the system in a quasi-equilibrium state is derived. A brief account of a unified consistent formalism associated with systems obeying power-law distributions precursor to the exponential form associated with thermodynamic equilibrium of systems is presented here.

1. **Introductory remarks**

The Boltzmann-Gibbs (BG) distribution has played a central role in explaining/understanding much of the equilibrium physical phenomena expressible in terms of macroscopic thermodynamic language [1, 2]. It is used without question in many circumstances and it works almost always. There are fundamental reasons for this success and usually we do not think of these reasons when BG is used. We therefore begin with a review of these basic foundations of BG. There are fundamentally two ways of understanding the foundations for the appearance of the BG distribution. One, based on probabilistic considerations providing the physical underpinning and, two, based on a maximization principle giving the variational foundation for the first, involving a functional quantity called "entropy". The introduction of entropy is central in obtaining the connection to thermodynamic description, e.g. the Legendre transformation structure. It should be noted that the BG distribution is of the exponential form which suits the thermodynamic description perfectly. It is, in the context of the second description, the idea of nonextensive/nonadditive generalization of the BG theory arose, because the power-law form of the distribution was found to be compelling for a cogent description a large number of observed complex phenomena such as turbulence, anomalous relaxation, multifractal structures, and many other facets of physical examples which are not in thermodynamic equilibrium. One of the first theoretical inquiries into the entropic origin of the BG distribution was raised by Jaynes [3] who asked "Why maximize entropy – why not some other measure of uncertainty?" Such a question was raised also in the context of the forms of nonextensive entropy [4] but we believe that we now have a better understanding of the answers to these questions. The origin of our subsequent work is a letter to us by Professor Balian in response to our publication [4] mentioned above, who pointed out to us that in [5], he with Balazs had established the uniqueness of the BG theory. In this rather personal presentation, we first give a careful analysis of both of these foundations of the BG distribution in order to develop a parallel set of arguments for a fundamental understanding of the nonextensive/nonadditive generalization. In particular, we show the uniqueness of the Tsallis distribution in describing phenomena requiring power-law distributions can be established on similar foundations.

## 2. Probabilistic foundations

Over the years, there are several probabilistic approaches to derive the BG distribution. We will present them here as concisely as possible and point out in each case the assumptions made therein. In the early stages the arguments were in terms of the thermodynamics language. But in recent years, these have been extended to include more general situations. Our presentation is thus ordered in increasing sophistication to cover many of these developments even though they are basically founded on similar fundamental principles.

### 2-a. Analysis of state density $\Omega$

In the first two subsections, we use the language of thermodynamics and statistical mechanics as in standard textbooks [1,2]. The macrostate of a system is described in terms of the number of possible microstates, $\Omega$. Consider two systems in thermal contact. Assume that they have fixed number of particles confined to their fixed volumes but their energies are variable with the constraint that their total energy is fixed. The actual form of this is not of concern in this analysis. Consider partitioning of energy between two systems while keeping the total energy fixed, ignoring interaction energy between them. Each of these subsystems are likely to be in any one of their own microstates $\Omega_i(E_i)$ ($i = 1, 2$), and therefore the composite of the two systems, $S_1$ and $S_2$, (obeying possibly different statistics as in Bose or Fermi or Maxwell) is just a product, $\Omega_1(E_1)\,\Omega_2(E_2)$ of the two numbers of microstates with the only condition that the sum of their energies is held fixed, $E_1 + E_2 = E$. The most probable partition of energy in conformity with the principle of equal *a priori* probability is found by using the mutual reciprocity of the logarithmic and exponential functions to show that the two attain the same "temperature" defined by $\beta \equiv (\partial \ln \Omega / \partial E)_{N,V}$. This defines the thermal equilibrium between the two systems.

Consider now system $S_1$ to be an objective system and immersed in the very large heat bath (reservoir) $S_2$: $E_1 \ll E_2$. In thermal equilibrium, they have the same temperature. Consider system $S_1$ to be in the $i$th state of energy $E_1 = \varepsilon_i$. The probability of finding it in such a state is proportional to the number of microstates of system $S_2$:

$$f(\varepsilon_i) \propto \frac{\Omega_2(E - \varepsilon_i)}{\Omega_2(E)} \qquad (1)$$

With the state of the system $S_1$ specified, the reservoir $S_2$ can still be in any one of a large number of states compatible with the energy value $E_2 = E - \varepsilon_i$. Since possible states with a given energy value are equally likely to occur, the probability in eq. (1) is computed using the mutual reciprocity of the logarithmic and exponential functions (to logarithmic accuracy since $\Omega_2$ is large):

$$f(\varepsilon_i) \propto \exp\{\ln\Omega_2(E - \varepsilon_i) - \ln\Omega_2(E)\} \propto \exp(-\beta\varepsilon_i). \qquad (2)$$

Note that this bears the relation to the reservoir $S_2$ only through $\beta$. This is the well-known Boltzmann factor. It should be remarked that this derivation is based entirely on macroscopic considerations and probability concepts.

We now consider the ensemble approach to the same problem, which takes a microscopic view of the same question.

2-b.    Ensemble weights theory

Consider an isolated system with energy $E$ consisting of $N$ non-interacting particles with states $\{i\}$ and energies $\{\varepsilon_i\}$. If $n_i$ is the number of particles in the $i$th state among the total $N$ particles, assuming all the states $\{\varepsilon_i\}$ are equally likely to be occupied, the probability of a configuration $\{n_i\}$ of the composite system is proportional to $W = N! / \prod_i n_i!$. The constraints on these numbers are that $\sum_i n_i = N$ and $\sum_i n_i \varepsilon_i = E$. (N.B. Fermi, Bose, and Maxwell statistics are all basically founded on "binomial" counting.) Then the most probable distribution, $p_i^{(0)} = n_i / N$, of finding the system in the $i$th state in the large $N$ limit (with the help of the usual maximization principle as well as the Stirling asymptotic formula) is given by $p_i^{(0)} = \exp(-\beta \varepsilon_i) / \sum_k \exp(-\beta \varepsilon_k)$. Here $\beta$ is determined by the energy constraint.

Now, if the states are occupied by a distribution $\{p_i^{(1)}\}$ not constrained as above, how far are these distributions from those obtained above, in the large $N$ limit? Consider then the configuration probability (apart from a normalization factor) $W(p^{(0)} | p^{(1)}) = W \times \prod_i (p_i^{(1)})^{n_i}$, where $W$ is as given above. This follows by the usual rule of computing probabilities. Again using the Stirling asymptotic formula, we obtain $W(p^{(0)} | p^{(1)}) \cong \exp\{-N \sum_i p_i^{(0)} \ln(p_i^{(0)} / p_i^{(1)})\}$. Taking $p_i^{(1)} = p_i^{(0)} + \Delta p_i$ and working to the leading order, we obtain $W(p^{(0)} | p^{(1)}) = A \exp\{-(N/2) \sum_i (\Delta p_i)^2 / p_i^{(0)}\}$, where $A$ is a normalization factor. For large $N$, this is very small unless $\Delta p_i$ is very small.

From this we see that the mean value of the occupation number, $n_i$, turns out to be $p_i^{(1)}$: $p_i^{(1)} = <n_i>\big|_{W(p^{(0)} | p^{(1)})}$. This procedure is then a "mean value" description of the same problem as opposed to the "most probable value" description and the two are thus shown to be equivalent in the large $N$ limit. The mean value theory is the basis of the general entropic methods as will be described later briefly [5].

Thus, for large $N$, the BG distribution is the unique limit and the deviation from it is due to the presence of the system-bath interactions.

Two crucial steps in the above description are as follows. $W$ chosen above is based on the binomial counting scheme. Also, in accommodating the constraints in finding $W$ we used mutual reciprocity of the logarithmic and exponential functions.

## 2-c. Steepest descents method

Here the counting is not explicitly binomial and the constraints of total number and total energy are taken into account quite generally, as in [5]. One still makes use of the equi-probability assumption and the idea of mutual reciprocity of the logarithmic and exponential functions enters in a different way.

The standard discussion [5] consists in making the supersystem $S = \{S_\alpha\}_{\alpha=1,2,\cdots,N}$ composed of a large number $N$ of replicas $S_1, S_2, \cdots, S_N$ of a classical system s. Let $A_\alpha$ be a physical quantity (such as the energy in the previous subsections) associated with the system $S_\alpha$. This is a statistical variable whose value is denoted by $a(i_\alpha)$ where $i_\alpha$ labels the allowed configurations of $S_\alpha$. $a(i_\alpha)$ is assumed to be bounded from below. The quantity of interest is the mean value of $\{A_\alpha\}_{\alpha=1,2,\cdots,N}$ over the system: $(1/N)\sum_{\alpha=1}^{N} A_\alpha$. The equi-probability assumption in the microcanonical ensemble theory requires that the probabilities of finding the supersystem S in the configurations in which the values of the mean value defined above lies around a given value $\bar{a}$, i.e.,

$$\left| \frac{1}{N} \sum_{\alpha=1}^{N} a(i_\alpha) - \bar{a} \right| < \frac{1}{N} \sum_{\alpha=1}^{N} \left| a(i_\alpha) - \bar{a} \right| < \varepsilon, \quad \varepsilon = O(1/\sqrt{N}) \tag{3}$$

are all the same. This equi-probability is given by

$$P(i_1, i_2, \cdots, i_N) \propto \theta(\varepsilon - |M|), \tag{4}$$

where $M \equiv (1/N)\sum_{\alpha=1}^{N} [a(i_\alpha) - \bar{a}]$ and $\theta(x)$ is the Heaviside unit step function. The probability of finding the system, say, $S_1$, in the configuration $i_1 = i$ is given by

$$p_i = \sum_{i_2, i_3, \cdots, i_N} P(i, i_2, i_3, \cdots, i_N), \tag{5}$$

which describes the canonical ensemble. Now the step function in eq. (4) has the following integral representation:

$$\theta(x) = \int_{\beta-i\infty}^{\beta+i\infty} d\phi \, \frac{e^{\phi x}}{2\pi i \phi} \tag{6}$$

with $\beta$ an arbitrary positive constant. Herein enters the exponential function in the theory. Further manipulation of the integral proceeds by the method of steepest descents to obtain the large $N$ limit, as in the earlier description. An important ingredient here is the property of the exponential function: $e^{x+y} = e^x e^y$. The final result is the well-known normalized BG distribution in this general context:

$$p_i = \exp[-\beta^*(a_i - \bar{a})]/Z(\beta^*), \quad \bar{a} = \sum_i p_i a_i \equiv <A>_1, \tag{7}$$

where $a_i \equiv a(i)$, $Z(\beta^*) = \sum_i \exp[-\beta^*(a_i - \bar{a})]$, with $\beta^*$ the steepest descent point determined by the mean value relation deduced above. Note that we did not use the binomial counting in this derivation. Note also that the arithmetic mean $\bar{a}$ is now expressed as the expectation value of $a_i$ over the probability distribution $p_i$.

We next turn to an equally brief account of the counting algorithm.

### 2-d. Counting algorithm

In this method [5], one employs a logarithmic counting algorithm. The probability $p_i$ of finding the objective system $S \equiv S_1$ in its $i$th configuration is given by the ratio of two numbers of equally-probable configurations $\{i_1, i_2, \cdots, i_N\}$, i.e., $p_i = W_i / W$. Here $W$ is the total number of configurations satisfying eq. (3), whereas $W_i$ is determined by the two conditions: (a) $S_1$ is found in $i_1 = i$, and (b) $\bar{a}$ in eq. (3) is given by the arithmetic mean over the configurations of S. This is done by noting that eq. (3) is rewritten in the form:

$$\left| \frac{1}{N}[a_i - \bar{a}] + \frac{1}{N} \sum_{\alpha=2}^{N} a(i_\alpha) - \frac{N-1}{N} \bar{a} \right| < \varepsilon. \tag{8}$$

The number of configurations $Y_N$ satisfying $\left|(1/N)\sum_{\alpha=2}^{N} a(i_\alpha) - (1 - 1/N)\bar{a}\right| < \varepsilon$ is counted in the large $N$ limit as follows: $\ln\left[Y_N((1-1/N)\bar{a})\right]^{1/N} \cong \ln\left[Y_N(\bar{a})\right]^{1/N} = S(\bar{a})$. From eq. (8) and the above condition, we have

$$\ln W_i = \ln Y_N\left(\frac{N-1}{N}\bar{a} - \frac{1}{N}(a_i - \bar{a})\right) \cong \ln Y_N(\bar{a}) - (a_i - \bar{a})\frac{\partial S}{\partial \bar{a}}. \tag{9}$$

Defining $\beta = \partial S(\bar{a})/\partial \bar{a}$, and putting $\tilde{Z}(\beta) = \lim_{N \to \infty} W / Y_N(\bar{a})$, we obtain the canonical distribution in the usual form after exponentiation of eq. (9). Thus we have used the property of mutual reciprocity of the logarithmic and exponential functions.

We note that the logarithmic counting was an essential step in this derivation because as in the beginning paragraphs, evaluation of limits of large numbers is best handled in this manner. Finally, we turn our attention to the method based on the central limit theorem and the law of large numbers.

### 2-e. Method based on central limit theorem

In the above description, we considered discrete physical entities. In this development, for the sake of simplicity of the presentation, we consider continuous quantities, the energy being a representative example, in order to exploit the limit theorems in probability theory. First recall a version of the central limit theorem, following [6]. In the spirit of what we have been discussing all along, given a system composed of a large number of subsystems with the rescaled energies, $\{\tilde{\varepsilon}_i = E_i / B_K\}_{i=1,2,\cdots,K}$, where $B_K$ is a positive $K$-dependent factor to be determined subsequently, let us consider the rescaled total energy $\tilde{E} = \tilde{\varepsilon}_1 + \tilde{\varepsilon}_2 + \cdots + \tilde{\varepsilon}_K$. The central limit theorem states that if each of $E_i$'s obeys a common distribution $f(E_i)$ with the ordinary finite second moment, $<E_i^2>_1 = \int_0^\infty dE_i\, E_i^2 f(E_i)$, then its $K$-fold convolution, $B_K (f * f * \cdots * f)(B_K \tilde{E})$, approaches the Gaussian distribution in the limit of large $K$, where $(f * g)(x) \equiv \int_0^x dx'\, f(x-x') g(x')$. This property was exploited by Khinchin [6] to establish the Gibbs canonical ensemble within the framework of probability theory. He employs the so-called generating function involving the Laplace transform of the number of microstates $\Omega$ introduced in 2-a. Note that the convolution property is fundamentally associated with the Laplace transform, which involves exponential function.

We now discuss briefly the method of maximum entropy leading to the BG distribution and expose the underlying assumptions in such formalism in the same fashion as was done above for the probabilistic formulation.

### 2-f. Maximum entropy method

The maximum entropy method in its most general form was given by Jaynes [3]. Stated in its simplest manner (which is sufficient for our purposes here), the entropy functional is of the form considered by Boltzmann, Gibbs, Shannon and Jaynes in different contexts. For a set of probabilities $\{p_i\}_{i=1,2,\cdots,W}$ of occurrences of $W$ events, it is given by

$$S[p] = -\sum_{i=1}^{W} p_i \ln p_i. \qquad (10)$$

Henceforth we call this the Boltzmann-Gibbs-Shannon entropy. The probabilities $\{p_i\}$ are determined by finding maximum of this functional if we are given constraints such as mean value of a physical quantity $A$, which takes on values $\{a_i\}$ upon its determination, defined by $\bar{a} = \sum_{i=1}^{W} p_i a_i$. Then we obtain the BG form for it: $p_i = \exp[-\lambda(a_i - \bar{a})]/Z(\lambda)$ and the parameter $\lambda$ is a Lagrange multiplier determined by the mean value constraint condition mentioned above and $Z$ is a factor that normalizes the total probability to be unity. If there is no constraint on the mean value, all the events are equally likely to occur: $p_i = 1/W$ for all $i$, and then the entropy is maximum. This derivation of the BG theory is based on the particular entropy given in eq. (10).

An important point to note at this juncture is that the Lagrange multiplier in this formulation is the same as the constants that appeared in the various forms of the derivations of the BG theory given above. This brings the probabilistic and entropic approaches together and relates them to the familiar thermodynamic framework.

There have been several axiomatic derivations of eq. (10) for entropy, notably those given by Shannon (in his context of information theory) and Khinchin (from mathematical viewpoint). Two of the requirements that are of interest to us here are (i) concavity and (ii) additivity. If $p_i^{(1)}$ and $p_i^{(2)}$ are two probabilities and $x$ is a parameter in the interval $(0,1)$, then the concavity of $S$ is expressed by the inequality $S[x p^{(1)} + (1-x) p^{(2)}] \geq x S[p^{(1)}] + (1-x) S[p^{(2)}]$. The additive property in its simplest form states that if two events are independent, the joint probability of their simultaneous occurrence is just the product of the probabilities, and then the total entropy of such an event is the sum of the two entropies. These two properties are important in deducing thermodynamics from statistical mechanical principles and in information theory.

It is useful to point out that Rényi [7] proposed his entropy functional:

$$S_\alpha^R[p] = \frac{1}{1-\alpha} \ln \sum_{i=1}^{W} (p_i)^\alpha \qquad (\alpha \neq 1). \tag{11}$$

This quantity has additivity. While it reduces to the form given in eq. (10) for $\alpha \to 1$, it is not concave for $\alpha > 1$. Most importantly, it has been pointed out in [8,9] that the only additive entropy which is "stable" under "observability criterion" is that given by eq. (10) and the Rényi entropy is ruled out as a viable alternate.

From this review of all the known methods to establish the BG distribution in classical statistical mechanics, the following basic ideas are found to form the central core: (A) equi-probability, (B) the large $N$ limit that involves use of the logarithmic and exponential functions in some form or other. From the entropy viewpoint, the two basic properties employed are concavity and additivity. These are essential in giving a coherent and internally consistent thermodynamic description of the system. Let us keep these fundamental features in mind when we go on to discuss possible alternate distributions that may be of different structure than the BG theory.

### 3. Non-Gibbsian distributions

There are a number of complex systems whose statistical properties at the quasi-equilibrium states are well described by non-exponential form, specifically the $q$-exponential distributions. The $q$-logarithmic and $q$-exponential functions are given by

$$\ln_q(x) \equiv \frac{x^{1-q}-1}{1-q}, \quad e_q(x) \equiv \begin{cases} [1+(1-q)x]^{1/(1-q)} & (1+(1-q)x > 0) \\ 0 & (1+(1-q)x \leq 0) \end{cases}, \quad (12)$$

respectively. Note that they are reciprocal of each other. Also, they converge to their ordinary counterparts when $q \to 1$. These functions have the following important properties: $\ln_q(xy) = \ln_q(x) + \ln_q(y) + (1-q)\ln_q(x)\ln_q(y)$ and $e_q(x)e_q(y) = e_q(x+y+(1-q)xy)$. For $q > 1$ and large negative $x$, the $q$-exponential function is of the power law. These are anomalous in the sense that they are not of the exponential form and do not follow from the principle of maximum Boltzmann-Gibbs-Shannon entropy in eq. (10) when some constraints are specified. But the maximum Tsallis entropy principle [10,11] with specification of a different form of the constraints is able to provide the formalism parallel to the conventional framework to obtain the above-mentioned $q$-exponential distributions, which is a power-law distribution for $q > 1$. Along with this, a consistent form of thermodynamics follows. A desire to understand the Tsallis framework of the *q*-exponential distribution in the same manner as described in the previous section is the focus of a set of papers [12–16]. Unless otherwise stated, the considerations pertain to power-law situations where $q > 1$. A brief but cogent account of this will be given now.

We first note the modifications needed in the probabilistic arguments listed above. After that, we address the corresponding changes needed to make in the entropy concept conforming to the probabilistic arguments. With these two steps, we outline a basis for understanding the Tsallis formalism on a footing similar to the BG one.

In going through the various ways of deducing the BG distribution, we found a clue to obtain the *q*-exponential distribution. It consists in keeping only the equi-probability assumption but replacing the ordinary logarithmic and exponential pair in the probabilistic framework given above from the subsections, from 2-a to 2-f, in the previous section. The replacement we employ is the *q*-logarithmic and its inverse *q*-exponential pair of functions. These go over to the conventional pair for $q \to 1$. In place of the replacement of the arithmetic mean by the mean value taken with respect to the distribution in the Gibbs theory, eq. (7), it is deduced that we must use the mean value taken with respect to the escort distribution, $P_i^{(q)}$:

$$<a>_q = \sum_{i=1}^{W} P_i^{(q)} a_i, \qquad P_i^{(q)} \equiv \frac{(p_i)^q}{\sum_{j=1}^{W}(p_j)^q}. \tag{13}$$

For $q \to 1$, this is just the usual mean value. This is important for making mean values to be convergent when the distribution has the power-law structure. This also means that we cannot be using binomial counting method and the Stirling asymptotic formula in this new scheme, because we are now dealing with fractals and such, which do not admit conventional counting in the Euclidean (phase) space. The central limit theorem and the law of large numbers used by Khinchin in his work to deduce the BG distribution are now replaced by the Lévy-type generalized central limit theorem and the generalized law of large numbers, respectively.

### 3-a.  Analysis of state density $\Omega$

Consider two systems in thermal contact keeping all other properties the same as before in 2-a. We use the relation (12) in finding the most probable partition of energy in conformity with the principle of equal *a priori* probability. We find that the two attain the same "temperature" $\beta_q = \beta / \Omega_q^{1-q}$, where $\beta \equiv \left(\partial \ln_q \Omega_q / \partial E\right)_{N,V}$. Here, $\Omega_q$ is the state density in the new scheme of counting. (N.B. The two systems may have different values of $q$, in general, but their corresponding temperatures will be equal.) This defines the new thermal equilibrium between the two systems [12].

Consider as before the case when $S_1$ is a small system immersed in the very large heat bath (reservoir) $S_2$. The arguments are not changed because they are general probabilistic statements:

$$f(\varepsilon_i) \propto \frac{\Omega_{q,2}(E - \varepsilon_i)}{\Omega_{q,2}(E)} \tag{1'}$$

As before, we use *equally likely* argument but use the mutual reciprocity of the $q$-logarithmic and the $q$-exponential functions (to $q$-logarithmic accuracy) and obtain

$$f(\varepsilon_i) \propto e_q\left(\ln_q \frac{\Omega_{q,2}(E - \varepsilon_i)}{\Omega_{q,2}(E)}\right) \approx e_q(-\beta_q \varepsilon_i). \tag{2'}$$

This is the new factor in place of the Boltzmann factor [12]. As before, this derivation is based entirely on macroscopic considerations and probability concepts. In this development, there is no restriction on the values of $q$.

### 3-b.  Ensemble weights theory

Since we do not (as of now) have an alternate to binomial counting, we have no corresponding derivation of this framework. However, in the next paragraph, we exhibit the more sophisticated version of this theory given in 3-d.

### 3-c.     Steepest descents method

We again make use of the equi-probability assumption and the idea of mutual reciprocity of the $q$-logarithmic and the $q$-exponential functions, which enter in a different way. Following [13] we take $q > 1$.

The standard discussion [5] as in 2-c is followed except change eq. (3) to

$$\left| \frac{1}{N} \sum_{\alpha=1}^{N} a(i_\alpha) - \bar{a} \right| < \frac{1}{N} \sum_{\alpha=1}^{N} \left| a(i_\alpha) - \bar{a} \right| < \varepsilon, \qquad \varepsilon = O(N^{-1-\delta}) \quad (\delta > 0) \tag{3'}$$

The dependence of $\varepsilon$ on $N$ is changed in anticipation of the expected result to be in conformity with the power law and the generalized law of large numbers.

As before, we assume that $a(i_\alpha)$ is bounded from below. The equi-probability is given by eq. (4). The probability of finding the objective system $S_1$ in the configuration $i_1 = i$ is given by eq. (5) that characterizes the canonical ensemble. Now the step function in eq. (4) is represented by an integral over the $q$-exponential function in this new version:

$$\theta(x) = \int_{\tilde{\beta} - i\infty}^{\tilde{\beta} + i\infty} d\phi \, \frac{e_q(\phi x)}{2\pi i \phi} \tag{6'}$$

with $\tilde{\beta}$ an arbitrary positive constant. Further manipulation of the integral proceeds by modification of the method of steepest descents incorporating the properties of the $q$-exponential function $e_q(\phi[\varepsilon - M]) \approx e_q(\phi \varepsilon) \, e_q(-\phi M)$ and further factorization of $e_q(-\phi M)$ factorized into the product over the supersystem to obtain the large-$N$ limit that we seek, as in the earlier description. The middle part of the inequality in eq.(3') is used in factorization of the $q$-exponential of the sum. All these steps are justified if the condition $\varepsilon = O(N^{-1-\delta})$ ($\delta > 0$) is taken into account. The final result is [13]

$$p_i = e_q(-\beta_q[a_i - \bar{a}])/Z_q(\beta_q), \qquad \bar{a} = \sum_i P_i^{(q)} a_i \equiv <A>_q. \qquad (7')$$

Here $a_i \equiv a(i)$ and $\beta_q$ is the steepest descent point determined by the new mean value relation with respect to the escort distribution deduced above. It is of interest to note that the original arithmetic mean $\bar{a}$ is expressed as the expectation value over the escort distribution defined in eq. (13).

### 3-d.  Counting algorithm

Here we follow [14] and employ $q$-logarithmic counting algorithm. In this sequel, we assume $q > 1$. The probability $p_i$ of finding the system $S_1$ in its $i$th configuration is, as before, given by the ratio of two numbers of equally probable configurations $\{i, i_2, \cdots, i_N\}$. The number of configurations $Y_{q,N}$ satisfying $\left|(1/N)\sum_{\alpha=2}^{N} a(i_\alpha) - (1 - 1/N)\bar{a}\right| < \varepsilon$ is counted in the large-$N$ limit as follows: $\ln_q\left[Y_{q,N}((1-1/N)\bar{a})\right]^{1/N} \cong \ln_q\left[Y_{q,N}(\bar{a})\right]^{1/N}$, which is now some new function $S_q(\bar{a})$ of $\bar{a}$. Setting $a(i_1) = a(i) \equiv a_i$, thus we have

$$\ln_q W_q = \ln_q Y_{q,N}\left(\frac{N-1}{N}\bar{a} - \frac{1}{N}(a_i - \bar{a})\right)$$

$$\cong \ln_q Y_{q,N}(\bar{a}) - \frac{1}{N}(a_i - \bar{a})\frac{\partial \ln_q Y_{q,N}(\bar{a})}{\partial \bar{a}}. \qquad (9')$$

Using the definition given in eq. (12), $S_q(\bar{a}) = \ln_q\left[Y_{q,N}(\bar{a})\right]^{1/N}$, $\beta = \partial S_q(\bar{a})/\partial \bar{a}$, $\beta_q = \beta\left[Y_{q,N}(\bar{a})\right]^{(q-1)/N}$ and the identity $\ln_q(x/y) = y^{q-1}[\ln_q(x) - \ln_q(y)]$, we obtain the results of the form given in eq. (7'). Note the appearance of the renormalized temperature, which turns out to be consistent with the Clausius and Carathéodory principles [17, 18].

### 3-e. Method based on generalized central limit theorem

We now follow [15]. Let us first recall a version of the Lévy-type generalized central limit theorem. In the spirit of what we have been discussing all along, given a system composed of a large number of subsystems with rescaled energies, $\{\tilde{\varepsilon}_i = E_i / B_K > 0\}_{i=1,2,\cdots,K}$, where $B_K$ is a positive $K$-dependent factor to be determined subsequently, let us consider the rescaled total energy $\tilde{E} = \tilde{\varepsilon}_1 + \tilde{\varepsilon}_2 + \cdots + \tilde{\varepsilon}_K$. The generalized central limit theorem states that if each of the $E_i$'s obeys a common distribution $f(E_i)$ with the ordinary moments of all orders, $<E_i^n>_1 = \int_0^\infty dE_i \, E_i^n f(E_i)$ ($n = 1, 2, 3, \cdots$) divergent, then its $K$-fold convolution, $B_K(f * f * \cdots * f)(B_K \tilde{E})$, may still approach a limiting distribution in the limit of large $K$. The limiting distribution is referred to as the Lévy stable distribution, $F_\gamma(E_i)$. The explicit form of this function is not known but since our variables are bounded from below and are in the positive half space, the characteristic function is known. It has the form

$$\chi_\gamma(t) = \int_0^\infty dE_i \, e^{itE_i} F_\gamma(E_i) = \exp\left\{-a|t|^\gamma \exp\left[i \operatorname{sgn}(t) \frac{\theta \pi}{2}\right]\right\}. \tag{14}$$

Here $a$ is a positive constant, $\gamma$ is the Lévy index lying in the range $(0, 1)$ due to positivity of our random variables, $\theta$ is a constant satisfying $|\theta| \leq \gamma$ and $\operatorname{sgn}(t) = t/|t|$. In [15,19] this was employed to show that the $q$-exponential distribution approaches upon $K$-fold convolution in the limit of infinite $K$ to the Lévy distribution. This is in complete parallel to the demonstration of Khinchin who showed that the BG distribution approaches the Gaussian distribution upon $K$-fold convolution in the limit of infinite $K$. The parameters of the Lévy distribution are then related to those of the Tsallis distribution: $q = 1 + 1/(\gamma + 1)$, $B_K = K^{1/\gamma}$. Also the first moment with respect to the escort distribution is found to be finite, even though the ordinary moment is divergent.

Note that in this development the condition $q > 1$ is appropriate for the power-law distribution, for which the ordinary moments diverge. Appearance of the convergent $q$-mean values is thus a natural consequence of the theory.

### 3-f. Maximum Tsallis entropy method

The maximum Tsallis entropy principle is as stated originally as follows. That is, maximize

$$S_q[p] = \frac{1}{1-q}\left[\sum_{i=1}^{W}(p_i)^q - 1\right], \qquad (15)$$

subject to the normalization of probabilities, $\sum_{i=1}^{W} p_i = 1$ with the constraint given by eq. (13). This gives the distribution in terms of the $q$-exponential function with the corresponding changes in the structure of the Lagrange multiplier associated with the constraint. The procedure outlined above reconciles with the result obtained by this maximization process. In contrast, the maximum Rényi entropy method under similar constraint conditions also yield $q$-exponential distribution even though it is additive.

The first point we note is that the additivity law for the conventional entropy is modified: $S_q[p^{(1)}p^{(2)}] = S_q[p^{(1)}] + S_q[p^{(2)}] + (1-q)S_q[p^{(1)}]S_q[p^{(2)}]$. Also $S_q[p]$ is concave for all $q > 0$. In [20], the formal arguments given by Shannon and Khinchin are thus changed to accommodate the nonadditive feature. The stability question also is revisited [9] and the Tsallis form is found to be stable in the same sense as the Boltzmann-Gibbs-Shannon entropy is stable. Thus it appears that the statistical mechanical description of many phenomena requiring power-law distributions is on the same type of footing as the BG exponential distribution.

An important point to note at this juncture is that the Lagrange multiplier in this formulation is *not* the same as the constants that appeared in the various forms of the derivations of Tsallis distribution given above. However the probabilistic and entropy approaches are brought together after a renormalization of the temperature and thus conforms to familiar thermodynamic framework [17,18,21]. The Legendre transform feature of the thermodynamics is recaptured in this way.

## 4. Concluding remarks

Summarizing, the modifications made in the two-prong foundations of the BG distribution are the use of $q$-exponential and $q$-logarithmic functions in the probabilistic part and the modified constraint in the maximum Tsallis entropy framework. The latter form of the constraint was originally introduced in [11] from other physical/mathematical considerations. For a different presentation of this development, one may refer to [13].

Several remarks may be in order as conclusions of this approach to non-BG distributions. The Tsallis form is perhaps one among many for which we were able to provide modifications of the foundations and yet preserving all the tenets of the statistical mechanics (variational character, the Legendre transform structure, etc.) and thermodynamics (Clausius and Carathéodory principles and other various laws). In particular, in [17,18] starting with the statistical mechanical expressions for entropy and $q$-mean value of the Hamiltonian, the thermodynamic definitions of temperature of nonextensive systems, adiabatic and isothermal processes, etc. are deduced. In [21], macroscopic thermodynamics based on composable nonextensive entropies in accord with Carathéodory's theorem is established. The connection to the thermodynamic concepts of work and quantity of heat can also be established in the quasi-static regime. In other words, all the traditional features are maintained as we go from quasi-equilibrium to final thermodynamic equilibrium. Is there some deep meaning to this remarkable feature?

## Acknowledgments

Besides acknowledging the Office of Naval Research for partial support of this work, AKR thanks the Organizers for inviting and providing him the full support to participate in this CNLS Workshop. Partial support of MSF is also gratefully acknowledged.